\documentclass[aps,prl,twocolumn,superscriptaddress,10pt]{revtex4-1}

\usepackage{graphicx}
\usepackage{amsmath}

\usepackage{color}

\begin{document}


\title{Widely tunable single-photon source from a carbon nanotube in the Purcell regime.}

\author{A. Jeantet}
\author{Y. Chassagneux}
\author{C. Raynaud}
\affiliation{Laboratoire Pierre Aigrain, \'Ecole Normale Sup\'erieure, CNRS, Universit\'e Pierre et Marie Curie, Universit\'e Paris Diderot, 24, rue Lhomond, F-75005 Paris, France}

\author{Ph. Roussignol}
\author{J.S. Lauret}
\affiliation{Laboratoire Aim\'e Cotton, CNRS, \'Ecole Normale Sup\'erieure de Cachan, Universite Paris Sud, 91405 Orsay, France}
\author{B. Besga}
\author{J. Est\`eve}
\author{J. Reichel}
\affiliation{Laboratoire Kastler Brossel, \'Ecole Normale Sup\'erieure, CNRS, Universit\'e Pierre et Marie Curie, 24 rue Lhomond, F-75005 Paris, France}

\author{C. Voisin}
\affiliation{Laboratoire Pierre Aigrain, \'Ecole Normale Sup\'erieure, CNRS, Universit\'e Pierre et Marie Curie, Universit\'e Paris Diderot, 24, rue Lhomond, F-75005 Paris, France}


\date{\today}

%
%


\maketitle



 \textbf{Single-Wall Carbon Nanotubes (SWNTs) are among the very few candidates for single-photon sources operating in the telecom bands since they exhibit large photon antibunching up to room temperature \cite{Hogele2008,Walden2012,Endo2015,Doorn2015,Ma2015}. 
 However, coupling a nanotube to a photonic structure is highly challenging because of the random location and emission wavelength in the growth process \cite{Xia2008,Watahiki2012,Legrand2013,Noury2014}. Here, we demonstrate the realization of a widely tunable single-photon source  by using a carbon nanotube inserted in an original repositionable fiber micro-cavity : we fully characterize the emitter in the free-space and subsequently form the cavity around the nanotube. This brings an invaluable insight into the emergence of quantum electrodynamical effects. We observe an efficient funneling of the emission into the cavity mode with a strong sub-Poissonian statistics together with an up to 6-fold Purcell enhancement factor. By exploiting the cavity feeding effect on the phonon wings, we locked the single-photon emission at the cavity frequency over a 4~THz-wide band while keeping the mode width below 80~GHz. This paves the way to multiplexing and multiple qubit coupling.}
 
Coupling a carbon nanotube to a photonic resonator in a reliable way is highly desirable both for technological developments in view of quantum cryptography or quantum computation and for academical studies since nanotubes behave like an original nano-emitter showing an hybrid 1D-0D electronic behavior \cite{Ardizzone2015}. In addition, the low-cost, the high integrability and the possible electrical excitation of nanotubes \cite{Marty2006} are attractive assets in such perspectives. However, due to the lack of control of the current growth or deposition processes, current attempts rely on random spectral and spatial matching between a resonator (micro-discs \cite{Imamura2013} or photonic crystals \cite{Miura2014}) and randomly deposited nanotubes, leading to a very limited fabrication yield. This constrain becomes especially stringent when a high coupling between the emitter and the cavity is sought, which requires narrow spectral features.  
 
 In this work, we propose an original approach where the nanotube is fully characterized in free-space by regular low-temperature micro-photoluminescence (micro-PL) spectroscopy and where a micro-cavity is subsequently formed around the emitter by approaching a concave dielectric mirror micro-engineered at the apex of an optical fiber. This geometry brings an unprecedented flexibility giving a built-in spectral and spatial matching, together with excellent quality factors and mode volumes \cite{Hunger2010}. Individual carbon nanotubes embedded in a polystyrene matrix were coupled to the cavity resulting in a strong brightening of the nanotube of more than an order of magnitude, bringing evidence for the relevance of exploiting cavity quantum electrodynamical (CQED) effects to enhance the photonic properties of carbon nanotubes. By means of time-resolved measurements we were able to investigate directly the cavity-enhanced emission rate and found a corresponding Purcell factor $F_p$ of up to 5. 
 In the same time, we performed intensity correlation measurements in a Hanbury-Brown and Twiss configuration and we checked that the photon flux retained a strong sub-Poissonian statistics with $g^{(2)}(0) \simeq 0.05$ both 
for free-space and cavity enhanced emissions. Finally, by adjusting the cavity length, we were able to tune the working wavelength of the emitter over 10~nm (4 THz) while keeping a high-spectral purity of the source (below 80 GHz) by exploiting the cavity feeding effect \cite{Auffeves2009} on the acoustic phonon side-bands of the luminescence of the nanotube \cite{Vialla2014}.

\begin{figure}
\includegraphics[width=8cm]{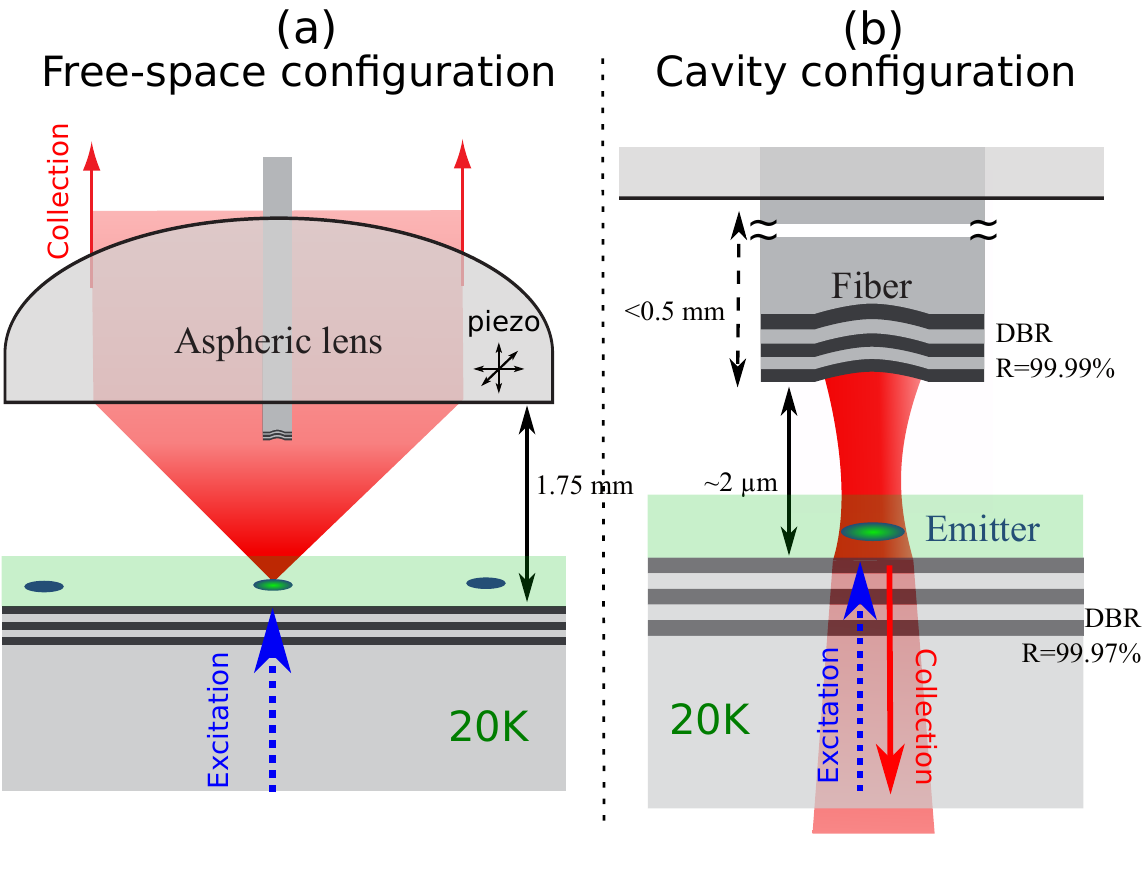}

\includegraphics[width=8cm]{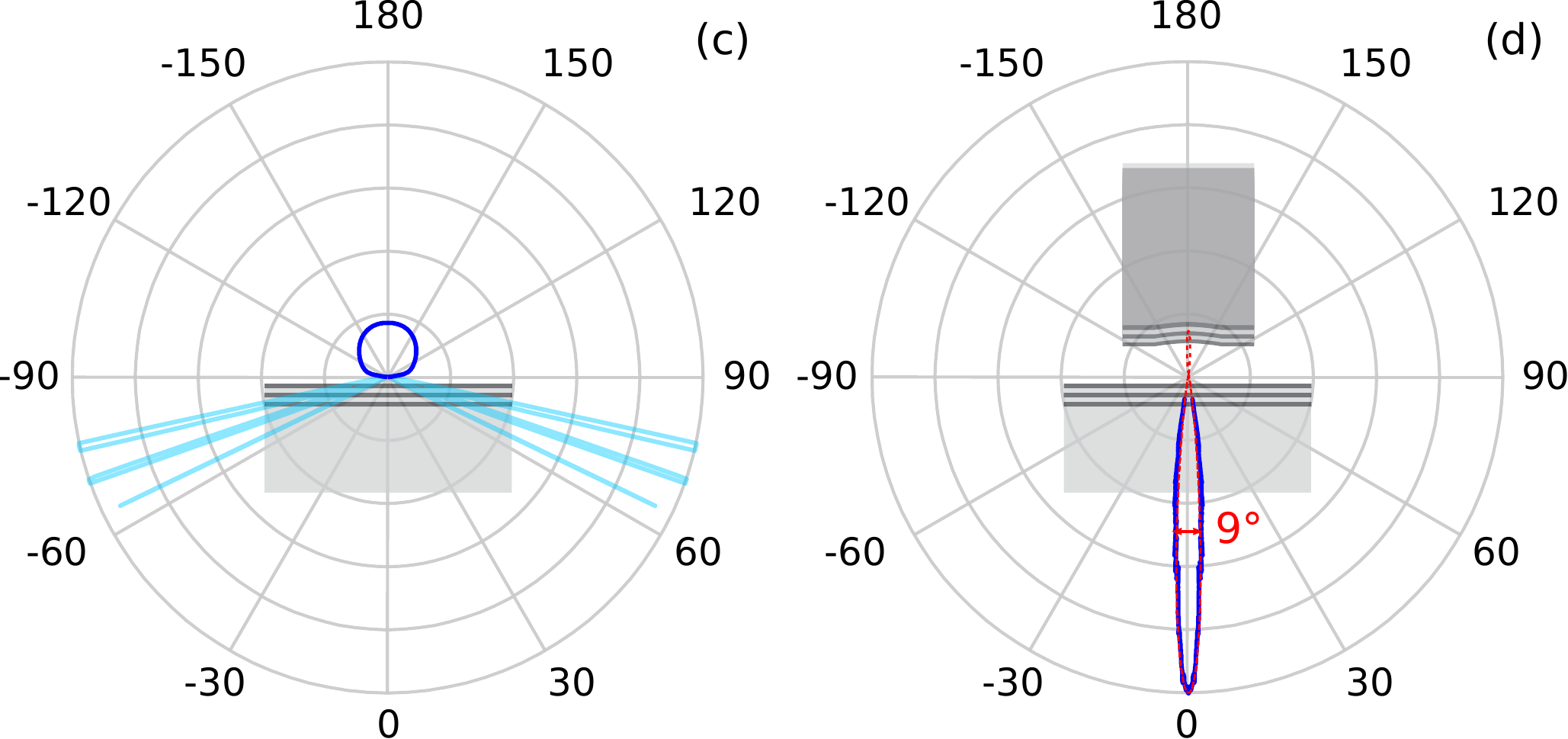}
\caption{(a)-(b) Sketch of the experimental setup showing the versatile micro-PL/scanning cavity microscope. The whole setup is placed on the cold finger of a helium cryostat at 20~K. (c) Calculated free-space radiation pattern of the nanotube placed at 20~nm of a reflective dielectric mirror. The light-blue pattern corresponds to the emission into the substrate which is lost. (d) Experimental radiation pattern of the nanotube resonantly coupled to the cavity as measured from a far field image of the beam (blue solid line); theoretical emission diagram calculated from the geometrical parameters of the cavity (see SI, red dotted line). The diagrams are normalized to the integrated emission. \label{fig:setup}}
\end{figure}

Figure~\ref{fig:setup} shows a sketch of the experimental setup. All the measurements were conducted at 20~K (see Methods for details). The sample was illuminated from the back with a cw or pulsed Ti:sapphire laser tuned beyond the edge of the stop-band of the mirror at about 800~nm corresponding to non-resonant excitation of the nanotube \cite{Vialla2014a}. An aspherical lens was used to collect the near-infrared luminescence of individual nanotubes. 
Once a nanotube was localized and characterized, the micro-cavity was formed by approaching the end-mirror of the cavity which is engineered at the apex of an optical fiber. In order to 
facilitate the positioning of the fiber with respect to the tube location, the fiber was inserted in the center of the aspherical lens used for the micro-PL measurements through a 400~$\mu$m hole.
Importantly, the excitation through the back mirror is unchanged in the cavity configuration, which ensures that the very same nanotube is investigated (see Methods). 
Finally, by choosing the reflectivities of the mirrors appropriately, we obtained an asymmetric Fabry-Perot cavity showing a preferential output (88\%) through the back mirror. In the cavity configuration, the output photons are thus collected from the back through a second aspherical lens and a dichroic mirror separating the infrared photon flux from the visible excitation beam. The finesse $F$ of the cavity was deliberately limited to about 6000 (at 900~nm) in order to match the quality factor of the emitter for an optimal coupling.

\begin{figure}
\includegraphics[width=8.5cm]{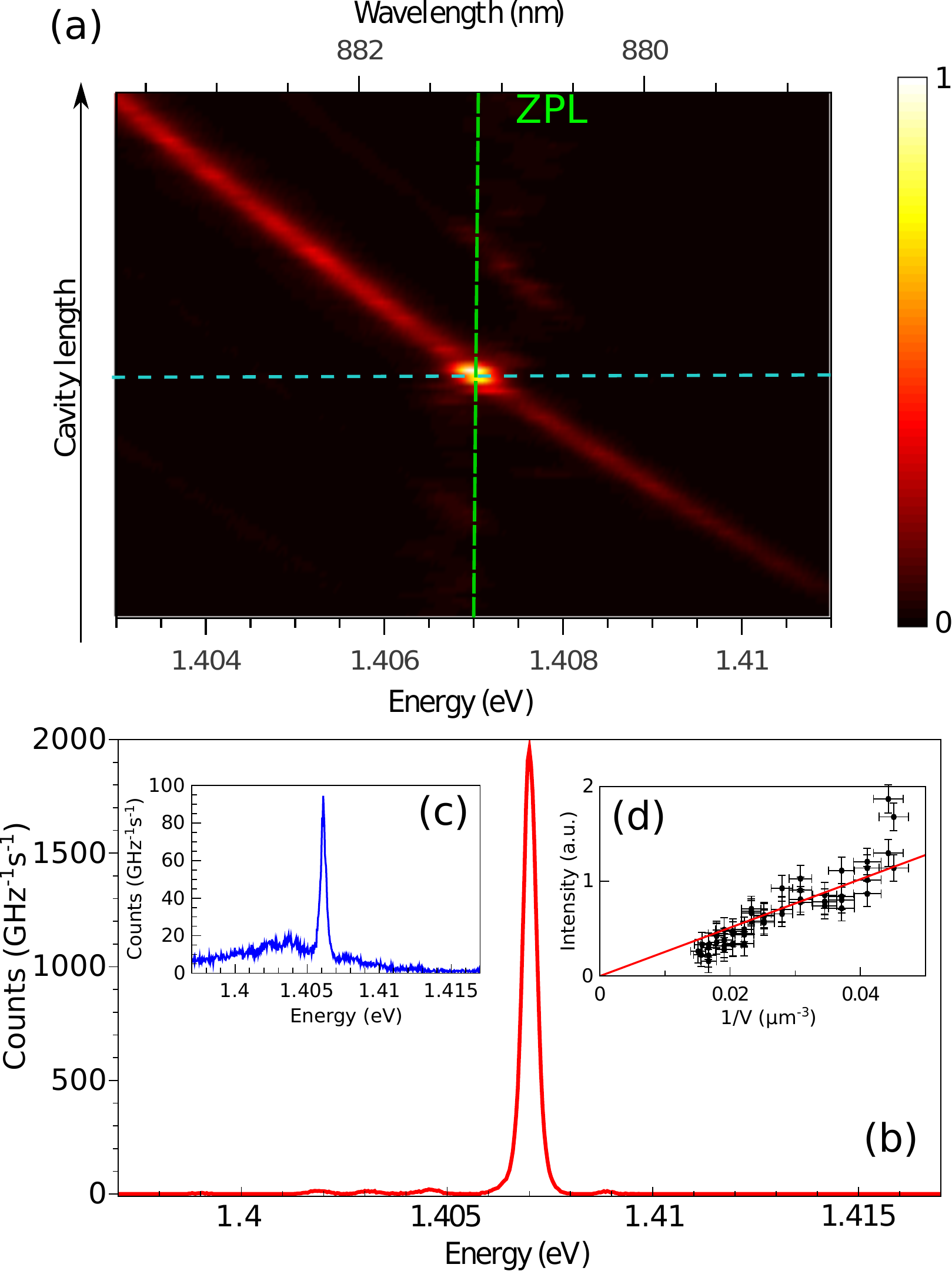}
\caption{(a) 2D plot of the emission spectrum of a nanotube embedded in the fiber micro-cavity. The color scale encodes the emission intensity while the horizontal axis corresponds to the wavelength and the vertical one to the cavity length. The green dashed line corresponds to the ZPL wavelength, while the light blue one corresponds to the resonant cavity length. (b) Emission spectrum (photons.GHz$^{-1}$.s$^{-1}$) of the nanotube when the cavity is tuned in resonance with its ZPL (red line, corresponding to the horizontal cut of panel (a) along the blue dashed line). (c) Free-space luminescence spectrum of the same nanotube. (d) Integrated count rate as a function of the mode volume when the cavity length is increased by steps of $\lambda/2$. \label{fig:FSvsCAV}}
\end{figure}

A typical luminescence spectrum of an individual nanotube as measured in the micro-PL configuration is shown in the left inset of Figure~\ref{fig:FSvsCAV}(c). The spectrum consists in a sharp (FWHM 500~$\mu$eV) zero-phonon line (ZPL) superimposed to acoustic phonon side-bands \cite{Vialla2014}. When closing the cavity, we observe a strong emission through the back mirror as long as the cavity length, and therefore the cavity resonant frequency, is tuned in resonance with the ZPL (Figure~\ref{fig:FSvsCAV}(a)). 
In this configuration, the luminescence collected from the cavity consists in a sharp single line (FWHM 330$\mu$eV ($\simeq$80~GHz)) showing a strong enhancement of the peak intensity as compared to the free-space configuration (Figure~\ref{fig:FSvsCAV}(b)). Obviously, the overall spectral width of the emission is much larger in the free-space configuration than in the cavity one. Therefore, the proper way to compare the data and to show that the cavity acts beyond simple spectral filtering is to show an enhancement of the photon spectral density \cite{Albrecht2013}.
 This figure of merit brings a direct insight into CQED effects even if the emitter's linewidth is much larger than that of the cavity. We find an enhancement of spectral density of the order of 20 for the particular tube of Figure~\ref{fig:FSvsCAV}. 

Let us first emphasize that the signal is collected through a highly reflective mirror, which is a direct proof of the coupling of the nanotube to the cavity mode, since in the absence of the fiber (top mirror) no signal can be detected through the back mirror. The 20-fold intensity enhancement results both from an intrinsic brightening of the emitter (Purcell effect) and from a better collection of the light emitted in the cavity mode. The collection effect can readily be seen in Figure~\ref{fig:setup} (d) where the experimental radiation 
pattern of a nanotube is presented. This pattern was measured from the far field image of the mode (see SI). It shows an angular aperture lower than 9 degrees in good agreement with the one expected for the TEM00 cavity mode, bringing additional evidence for the good coupling of the emitter to the cavity. This narrow emission angle allows an almost 100\% collection efficiency of the photons emitted from the back mirror, in contrast to the free-space configuration (Figure~\ref{fig:setup} (c)) that shows a classical dipolar radiation pattern (modulated by interferences due to the back supporting mirror) leading to a collection efficiency of the order of 20\% only.


Taking into account these collection efficiencies and the transmission factors of all the optical parts of the setup, we can obtain a first estimate of the intrinsic brightening of the nanotube, that is the enhancement of its radiative rate \footnote{Broadband excitation was used in this measurement to avoid interferences effects in the excitation.}. In fact, as we shall see below, carbon nanotubes have a very low radiative yield and any increase of the radiative rate directly translates in an equivalent increase of the brightness, even for an excitation well below the saturation. As a consequence, exploiting CQED effects to enhance the brightness of a single photon source is especially relevant for such dim emitters. We can finally compare the spectrally integrated photon counts in both configurations and we find an intrinsic brightening of the order of 3. 

We now take advantage of the unique flexibility of this geometry to tune the length of the cavity over a wide range so as to change significantly the mode volume. The corresponding count rate is shown in Figure~\ref{fig:FSvsCAV}(d) for multiple back and forth scans of the cavity length. The length of the cavity is increased by step of $\lambda/2$ in order to remain in resonance with the ZPL of the emitter. The linear $1/V$ dependence of the count rate is directly related to the variation of the Purcell effect with the mode volume (see Eq.~\ref{eq:Purcell}).

\begin{figure}
 \includegraphics[width=7cm]{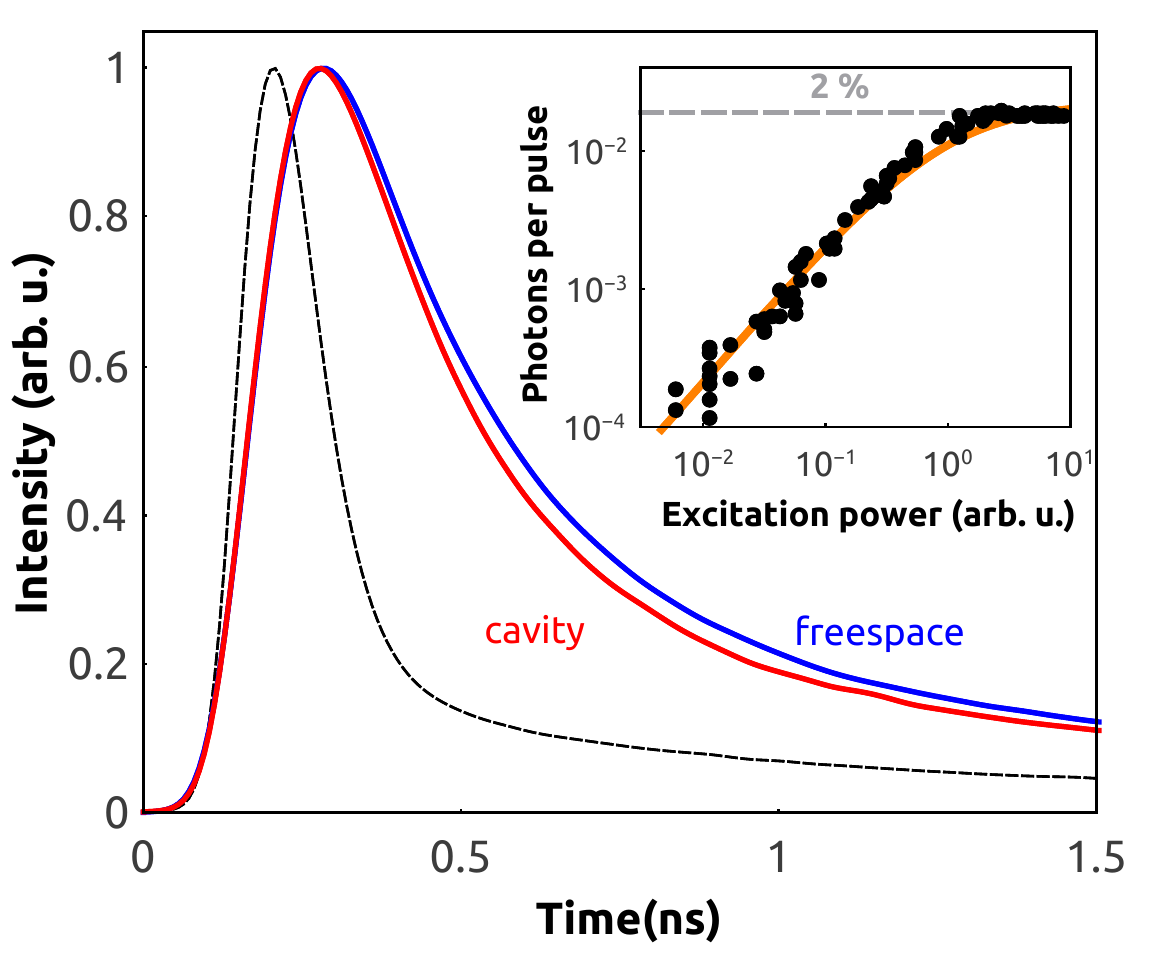}
 \caption{Time-resolved photoluminescence signal of a nanotube measured in the free-space configuration (blue line) or coupled to a resonant fiber cavity (red line) showing a 10\% reduction of the lifetime. Response of the detector (black line). The free-space transient has been convoluted with the empty cavity response (non shown) in order to account for the photon storage time in the cavity. Inset : Saturation measurement of the same nanotube measured in the free-space configuration under pulsed excitation. \label{fig:TR-PL}}
\end{figure}

Nevertheless, the most direct insight into the Purcell effect is obtained in the time domain, where the enhancement of the radiative rate can be directly measured \cite{Gerard1998}, by means of time-resolved photo-luminescence. In the case of a primarily non-radiative emitter however, the global lifetime is marginally modified. More precisely, the Purcell factor $F_p$ can be inferred from the relative change of lifetime through :
\begin{equation}
\tau_{cav}=\frac{\tau_{fs}}{1+\eta F_p}
\label{eq:taucav}
\end{equation}
where $\tau_{cav}$ is the lifetime of the nanotube in the cavity, $\tau_{fs}$ is its lifetime in free-space and $\eta$ is its quantum radiative efficiency (\textit{i.e.} radiative recombination rate over global transition rate). Figure~\ref{fig:TR-PL} shows time-resolved photoluminescence transients of a nanotube with and without the cavity. In order to account for the photon storage in the cavity, we separately recorded the decay of the cavity subsequent to a resonant pulsed excitation (see SI). This photon storage time turns out to be of the order of 35~ps. We convoluted this photon storage decay to the free-space decay of the nanotube in order to properly compare the life-times in either configuration. Using this procedure, we can make a direct comparison between the luminescence decays in the free-space and in the cavity configuration. Figure~\ref{fig:TR-PL} unambiguously shows an reduction of lifetime of the emitter when coupled to the cavity, which is a direct signature of the Purcell effect. The numerical fits of these transients yield $\tau_{cav}=183\pm4$~ps, $\tau_{fs}=203\pm 4$~ps. Thus, we deduce that the relative change of lifetime induced by the cavity is of the order of 10\%. Similar results were obtained on all the tubes we investigated, with a systematic cavity-induced decrease of lifetime (see SI).

The radiative quantum yield of this emitter was estimated through saturation measurements in pulsed excitation (Figure~\ref{fig:TR-PL}). In fact, in the saturation regime, one and only one 
excitation is created in the nanotube for each pulse (see intensity correlation measurements below) . Therefore, the count rate divided by the repetition rate of the laser is a direct measurement of the radiative yield of the nanotube. In the particular case of the nanotube of Figure~\ref{fig:TR-PL}, we find $\eta \simeq 2\%$. This value is consistent with previous estimates reported in the literature \cite{Berger2007}. 

Importantly, note that due to the dispersion of the radiative yield and lifetimes measured from tube to tube, the flexibility of the reconfigurable fibered cavity which allows to study the very same nanotube in free-space and in the cavity, turns out to be essential in order to obtain a reliable estimate of the magnitude of the CQED effects.

In total, these time-resolved measurements bring the first direct demonstration of a sizable Purcell effect with carbon nanotubes, with a Purcell factor $F_p$ of the order of 5 in the particular case of the nanotube shown in Figure~\ref{fig:TR-PL}. This value corresponds to a 6-fold enhancement of the radiative yield of the nanotube, bringing its value close to 12\%. Furthermore, this enhanced emission rate in the cavity mode corresponds to a coupling factor $\beta=F_p/(1+F_p)$ of the order of 0.8 meaning that 80\% of the photons emitted by the nanotube are effectively funneled into the cavity mode.

This $F_p$ value is in good agreement with the theoretical value (assuming a nanotube placed at a field maximum) given by \cite{Purcell1946}~:
\begin{equation}
\label{eq:Purcell}
F_p=\frac{3}{4 \pi^2} \frac{(\lambda/n)^3}{V}Q_{eff}
\end{equation}
 where $V$ is the mode volume, $n$ is the optical index at the position of the emitter and $Q_{eff}$ the effective quality factor of the system ($\frac{1}{Q_{eff}}= \frac{1}{Q_{cav}} + \frac{1}{Q_{em}}$). The cavity length (and thus the mode volume) can be extracted from the free spectral range (see SI), the effective quality factor is deduced from the white-light transmission measurements for the cavity contribution $Q_{cav}$ while the dephasing rate $\gamma^*$ of the emitter (and thus $Q_{em}=\frac{\hbar \omega_{ex}}{\gamma^*}$) is measured from the FWHM of the ZPL (500~$\mu$eV). We obtain $F_p^{th} = 5$, in good agreement with the experimental estimate. Interestingly, the strength of the coupling of the emitter to the confined optical mode can also be expressed in terms of the cavity-emitter coupling factor (or vacuum Rabi splitting) $g$, which can be compared to the other energy scales of the system. Using $F_p=4 g^2/\gamma_R \gamma^*$, where $\gamma_R$ stands for the radiative lifetime, and  
eq. (\ref{eq:taucav}), we deduce $g=(\hbar/2)\sqrt{\gamma^*(\tau_{fs}-\tau_{cav})/(\tau_{cav} \tau_{fs})}$. We obtain $g \sim 7 \, \mu$eV for the nanotube of Figure~\ref{fig:TR-PL}. 

We performed such combined time-resolved and saturation measurements on about a dozen of nanotubes. We consistently find a relative reduction of life-time of the order of 10\%, while the radiative yield spans the [0.5-5]\% range, leading to Purcell factor between 1 and 15.  This dispersion is not surprising since the radiative yield is known to be highly sensitive to defects and environment.  Similarly, we find a coupling $g$ ranging from 7 to 30 $\mu$eV. Note that the best values of $g$ lie only within a factor 5 below the line-width $\gamma^*$ of the narrowest nanotubes observed in this study, putting the strong-coupling at reach for reasonable improvements of the system.

\begin{figure}
 \includegraphics[width=9cm]{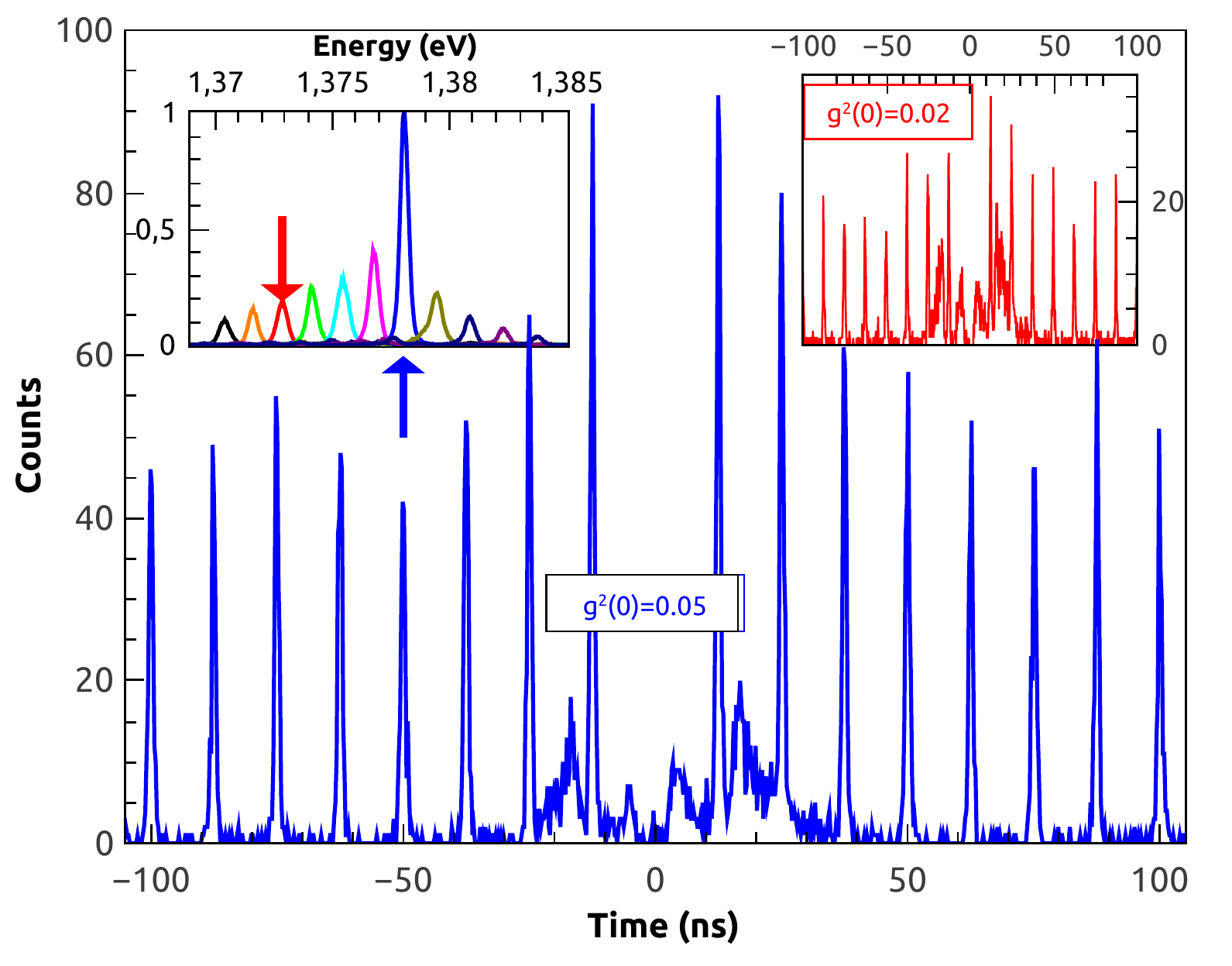}
 \caption{Intensity correlation measurements of a nanotube resonantly coupled to a fiber cavity obtained in a Hanbury-Brown and Twiss setup under pulsed excitation. The missing peak at time 0 ($g^{(2)}(0)=0.05$) shows that high-quality single photons are emitted by the nanotube. The signal between the first and second peaks correspond to detector artefacts (afterpulses). Inset : same measurements for a detuning of the cavity corresponding to an emission of the phonon side-wings. \label{fig:g2}}
\end{figure}

We now investigate the quantum properties of the source made of the nanotube resonantly coupled to the cavity by measuring the intensity correlations of the emitted light in Hanbury-Brown and Twiss setup (Figure~\ref{fig:g2}) \cite{Michler2000a}. Free-space emission of carbon nanotubes at low temperature is known to show strong antibunching \cite{Hogele2008}. However, the microscopic mechanism responsible for such photon statistics in a one-dimensional emitter is still a debated issue \cite{Hogele2008,Ma2015}. Despite the strong spectral filtering of the emission in the cavity configuration, we observe that this peculiar photon statistics is preserved with an almost full suppression of multiple photon emission probability ($g^{(2)}(0) \sim$ 0.05). This paves the way to a possible use of cavity embedded nanotubes as efficient near infrared single photon sources in the telecom bands -where alternative emitters are scarce- since telecom wavelengths are easily reached with slightly larger diameter nanotubes \cite{Ardizzone2015} (1.1~nm \emph{vs} 0.7~nm for the SWNTs used in this study).

Interestingly, the emission enhancement described before is not limited to the ZPL spectral range and we can make use of the phonon side-bands to efficiently feed the cavity in a spectral window as large as 15 meV (4~THz) around the ZPL. This effect is shown in the left inset of Figure~\ref{fig:g2} where the spectrum of the light collected from the cavity is displayed for several detunings, within the same longitudinal mode. This shows the ability of carbon nanotubes to be used as widely tunable single-photon sources. In fact, as can be seen in the upper right inset of Fig.~\ref{fig:g2}, the source retains its ability to emit single photons all throughout the 15~meV range of tunability. 
The origin of this efficient cavity feeding is not completely clear yet, but it could be related to the nature of the acoustic phonon wings. In fact, these wings imply long-lived acoustic phonons that do not introduce additional dephasing (in contrast to optical phonon replica \cite{Albrecht2013}). Therefore, the coupling of the wings to the optical mode is expected to be comparable to the one of the ZPL \cite{Hohenester2009}.
 
%

In conclusion, we have demonstrated that single-wall carbon nanotubes can be efficiently coupled to a micro-cavity to reach an up to six fold brightness enhancement. Using an original approach were the micro-cavity can be closed on a specific emitter after a full free-space characterization, we were able to perform a thorough investigation of the Purcell increase of the radiative rate due to the optical confinement. Finally, we have shown that the broad acoustic phonon sidebands in the luminescence spectra of carbon nanotubes can be exploited to feed the cavity over a broad spectral range, providing an original means to implement a narrow-band widely tunable single photon source. This pioneering study paves the way to many developments in exploiting CQED effects with carbon nanotubes by using the improvement margins of both the nanotube and the cavity. For instance, the use of slightly larger diameter nanotubes could provide a unique way to obtain an efficient single photon source operating at telecom 
wavelengths, whereas an improvement of the mirror curvature at the fiber apex, and therefore a reduction of the mode volume should put the strong coupling regime at reach, opening the way to hybrid excitations with a marked one-dimensional component.

\section*{Methods}

\paragraph{Sample}

CoMoCat nanotubes wrapped in PFO are dispersed in toluene with polystyrene
and spinned on the surface of a dielectric mirror resulting in a 120~nm
thick layer. The nanotube concentration is lower than 0.1~$\mu$m$^{-2}$.

\paragraph{Free-space PL }

Photoluminescence measurements were carried out in a closed-cycle
cryostat at 20~K using a home-made scanning confocal
microscope. SWNTs were excited with a 800~nm continuous
wave Ti:Sa laser, polarized along the SWNT axis, through the back mirror
{[}Fig 1{]} (reflectivity below 5\% at 800~nm) with
 an aspherical lens L1 ($NA=0.5$, spot size 70~$\mu$m$^{2}$). The luminescence was collected through a second aspherical lens L2 ($NA=0.68$) and was detected with a liquid nitrogen cooled silicon CCD camera coupled to a 500~mm grating spectrometer.

\paragraph{Cavity PL}

The Fabry-Perot cavity is composed of the plane back mirror on which
the tubes are deposited and a concave mirror (radius of curvature $\simeq$ 50$\mu$m
machined by laser ablation with a CO$_2$ laser at the apex of a multimode optical fiber inserted
in lens L2. The excitation is exactly similar to free-space PL configuration to ensure that the very same nanotube is excited. Due to the slight asymmetry in the reflectance of the mirrors, the emission is mostly (88\%) directed through the back mirror. It is further collected
by lens L1 and detected by the same spectrometer. Interferences on the excitation beam as a function of the cavity length were calibrated against a broad-band excitation with a super-continuum laser.

\paragraph{Lifetime Measurements}

Lifetime measurements were performed by exciting the nanotube with a femtosecond pulsed
Ti:Sa laser at 800~nm and 150~fs pulse length) and recorded with an avalanche photodiode
with a 40~ps time resolution. Laser
stray light was rejected by a 10~nm band-pass
filter. 
The photon storage time in the cavity is measured by generating a laser pulse at the emission wavelength of the nanotube with a nonlinear fiber.
This pulse is injected in the cavity formed at the exact location of the nanotube to ensure that the damping of the cavity is equal to the one involved in the luminescence measurements. The photoluminescence transients are fitted with the convolution of the experimental response function of the detector with a mono-exponential decay (free-space case) or with the convolution of the experimental photon storage decay of the cavity with a mono-exponential. 
Additional details are given in the SI.

\paragraph{Photon Correlations}

The intensity correlation measurements were carried out with a regular Hanburry-Brown
and Twiss setup. The emitter was excited as for lifetime measurements, while
the emission was filtered using the spectrometer as a monochromator.
The correlations were detected with a pair of 400~ps time-resolution
avalanche photodiodes.

\subsection*{Acknowledgement}
This work was supported by the C'Nano IdF through the grant \textit{"C'Nano IdF ECOQ"}. CV and JR are members of ``Institut Universitaire de France''.  We thank C. Vaneph for helpful discussions and N. Garroum for the mechanical fabrication, N. Izard for helping in sample preparation. \\

\subsection*{Authors contributions}
CV, YC and JR conceived the experiment. YC, AJ and CR designed the setup and collected the data. BB fabricated the fiber mirror. 
YC, AJ, CR, CV and JE analyzed the data. All the authors contributed to the writing of the paper.

\subsection*{Financial Interests}
The authors declare no competing financial interests.



\bibliography{purcell-NTcav}

\end{document}